# Dozer: Migrating Shell Commands to Ansible Modules via Execution Profiling and Synthesis

Eric Horton
ewhorton@ncsu.edu
North Carolina State University
Raleigh, NC, USA

Chris Parnin
cjparnin@ncsu.edu
North Carolina State University
Raleigh, NC, USA

## ABSTRACT

Software developers frequently use the system shell to perform configuration management tasks. Unfortunately, the shell does not scale well to large systems, and configuration management tools like Ansible are more difficult to learn. We address this problem with Dozer, a technique to help developers push their shell commands into Ansible task definitions. It operates by tracing and comparing system calls to find Ansible modules with similar behaviors to shell commands, then generating and validating migrations to find the task which produces the most similar changes to the system. Dozer is syntax agnostic, which should allow it to generalize to other configuration management platforms. We evaluate Dozer using datasets from open source configuration scripts.

## 1 DOZER

Using Bash scripts to manage infrastructure is, according to Netflix engineer Lorin Hochstein, "like the dark side of the force: quicker, easier, and more seductive, but not the right way to go" [7]. Despite this, developers frequently begin their configuration management journey with the shell because it is a familiar environment that provides them with a "quick and dirty" solution to their configuration management needs [4, 12]. Many of these developers will eventually discover that the shell is not without its growing pains and seek to integrate a full configuration management system like Ansible [2, 3, 8]. This happened to NASA when they migrated 65 legacy applications to the cloud and realized that their shell-based process made it difficult to do seemingly simple tasks like managing user accounts [5].

Dozer[1] is designed to address the often recommended approach of converting an existing shell script into an Ansible playbook: manually, one command at a time [1, 6, 9, 10]. It does so by accepting a shell command and returning an Ansible task that makes similar configuration changes to the system. Dozer is based on the insight that shell commands and Ansible modules can only change the system state by communicating with the kernel via the system call (syscall) interface. This shared interface provides an opportunity to observe and compare the behavior of different executables.

Before the Dozer migration pipeline begins, we collect system call traces (straces) for many Ansible module executions. These execution definitions and straces form the core of the Dozer knowledge base. When a developer wishes to migrate a shell command to Ansible, Dozer will compare the strace of the shell command to those in its knowledge base to find a collection of Ansible modules with similar behavior. We use a comparison scheme that attempts to map shell command parameters to Ansible module parameters for the best overall match, since recorded executions in the knowledge base likely used different parameters than the command to migrate, and weights syscall matches based on their information content to emphasize more infrequent syscalls [11]. Figure 1 gives a high-level view of how two straces are compared. Finally, Dozer uses the similar modules and parameter mappings discovered during the comparison step to generate different migrations of the shell command into an Ansible module. The shell command and each migration are executed against the same Docker image, and the migration with the most similar resulting system changes is selected.

## 2 EVALUATION

We evaluated Dozer on its ability to migrate 62 common shell commands found in open source Dockerfiles to Ansible modules. The Dockerfiles were sourced from projects on GitHub. Ansible modules were traced using the DebOps suite.[2] Migrations were assessed on Dozer's ability to select the correct Ansible module, select the correct module parameters, and to correctly map source to target parameters if applicable.

Overall, Dozer successfully chose the correct target module and parameters for 38 of the 62 commands being migrated. Figure 2 shows an example of one such migration of an echo shell command into an Ansible module (Figure 2a). Dozer first finds the definition of an Ansible module with similar behavior by comparing the strace of the shell command to the recorded straces of Ansible modules in its knowledge base. The similar module and inferred parameter mapping are used to generate the final migration to Ansible's

---

[1] https://github.com/config-migration/dozer
[2] https://docs.debops.org/en/stable-1.2/

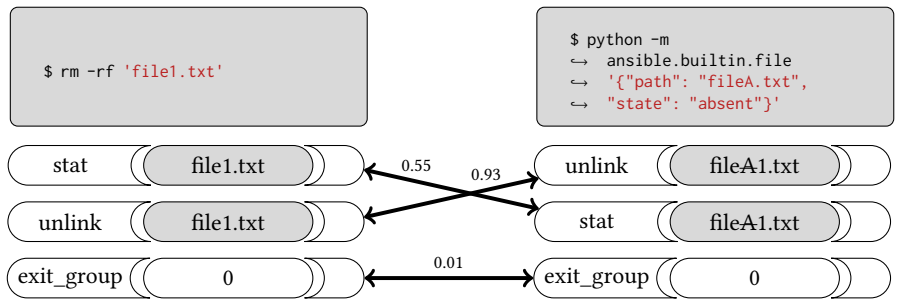

Figure 1: A high-level depiction of the strace comparison between the shell command rm and the Ansible module file. Dozer first searches for instances of parameters within syscalls, then determines the mapping that will result in the best score. Finally, it matches equivalent syscalls between the straces and assigns an overall comparison score based on the weighted scores of the matched syscalls.

```
echo 'daemon off;' >> /etc/nginx/nginx.conf
```

(a) An echo shell command that writes a configuration value to a line in a file.

```
lineinfile:
  dest:   '/root/.profile'
  regexp: '^.*mesg n.*$'
  line:   'tty -s && mesg n || true'
  state:  'present'
```

(b) An Ansible module with detected similar behavior to the shell command in Figure 2a.

```
lineinfile:
  dest:   '/etc/nginx/nginx.conf'
  regexp: '^.*mesg n.*$'
  line:   'daemon off;'
  state:  'present'
```

(c) Dozer's final migration of the shell command in Figure 2a into an Ansible module.

Figure 2

`lineinfile` module (Figure 2c). Note that `lineinfile` will append to the end of the file if its `regexp` parameter is not matched, so the final migration has the same effect on the system in a clean starting environment.

Some unsuccessful migrations were a result of Ansible not having a module that directly supported the same behavior as the shell command or were a result of the correct module not appearing in Dozer's knowledge base because it was not used in the DebOps suite we traced. In other cases, unsuccessful migrations resulted from Dozer being unable to detect similar behavior. For example, the shell command `mkdir -p <path>` splits the path argument into its component parts while the Ansible `file` module uses absolute paths. This mismatch adversely affects Dozer's comparison procedure.

## 3 DISCUSSION

Dozer presents a novel approach to migrating individual configuration tasks. We believe that this is a critical first step towards being able to migrate entire configuration scripts, since modern configuration management languages like Ansible, Puppet, Chef, etc. are composed of individual building blocks (Ansible modules, Puppet/Chef resources, …) that operate at approximately the same level of scope. However, there are additional challenges that need to be addressed in order to scale up to full configuration scripts. Dozer works by profiling a program's behavior based on its interaction with the syscall interface, allowing it to operate without an explicit domain knowledge of the underlying configuration script. This approach collects very little information about the system itself or the changes being made (outside of the final validation for similarity). We expect that migrations could be improved by incorporating additional information about changes to the system state into the search process.

Additional work is needed to solve the problem of composing configuration tasks, which is necessary when a task in one configuration language is equivalent to a sequence of two or more tasks in another. Notable challenges with composition include the correct propagation of information as outputs and inputs, selecting tasks that are "compatible" such that they don't conflict with each other or overwrite desired changes, and preserving control flow and error handling.

## ACKNOWLEDGMENTS
This work is funded in part by the NSF SHF grant #1814798.

## REFERENCES

[1] 2018. Bash scripts to Ansible. https://www.reddit.com/r/ansible/comments/a1qpr0/bash_scripts_to_ansible/.
[2] 2019. https://news.ycombinator.com/item?id=20379217.
[3] 2020. https://news.ycombinator.com/item?id=20375380.
[4] 2020. Ansible versus BASH script. https://www.reddit.com/r/linuxadmin/comments/emcuqm/ansible_versus_bash_script/.
[5] Ansible. 2016. NASA: INCREASING CLOUD EFFICIENCY WITH ANSIBLE AND ANSIBLE TOWER. https://www.ansible.com/hs-fs/hub/330046/file-1649288715-pdf/Whitepapers__Case_Studies/nasa_ansible_case_study.pdf.
[6] Allen Eastwood. 2018. Shell Scripts to Ansible. https://www.ansible.com/blog/shell-scripts-to-ansible.
[7] Lorin Hochstein. 2015. https://twitter.com/norootcause/status/679731676193230849.
[8] Jonah Horowitz. 2017. Configuration Management is an Antipattern. https://hackernoon.com/configuration-management-is-an-antipattern-e677e34be64c.
[9] Matt Jaynes. 2013. Shell Scripts vs Ansible: Fight! https://hvops.com/articles/ansible-vs-shell-scripts/.
[10] Luke Rawlins. 2018. How to get started using Ansible. https://sudoedit.com/how-to-get-started-using-ansible/.
[11] C. E. Shannon. 1948. A mathematical theory of communication. *The Bell System Technical Journal* 27, 3 (1948), 379–423.
[12] Aaron Weiss, Arjun Guha, and Yuriy Brun. 2017. Tortoise: Interactive System Configuration Repair. In *Proceedings of the 32Nd IEEE/ACM International Conference on Automated Software Engineering* (Urbana-Champaign, IL, USA) *(ASE 2017)*. IEEE Press, Piscataway, NJ, USA, 625–636. http://dl.acm.org/citation.cfm?id=3155562.3155641